\documentclass{article}
\usepackage{float}
\usepackage{authblk}

\title{Equivalence of Singles-server and Multiple-servers Blind Quantum Computation Protocols}
\author[1]{Yuichi Sano\thanks{\texttt{sano.yuichi.77v@st.kyoto-u.ac.jp}}} 
\affil[1]{Department of Nuclear Engineering, Kyoto University, Nishikyo-ku, Kyoto 615-8540, Japan} 

\date{\today}

\usepackage{graphicx}
\usepackage{amsmath,amsthm,amssymb}
\usepackage{amsfonts}
\usepackage{siunitx}

\usepackage{qcircuit}
\usepackage{braket}
\usepackage{cite}
\usepackage{comment}

\theoremstyle{definition}
\newtheorem{thm}{Theorem}
\newtheorem{defn}{Definition}

\begin{document}

\maketitle

\begin{abstract}
Because quantum computers are expensive, it is envisaged that individuals who want to utilize them would do so by delegating their calculations to someone who has a quantum computer.
When quantum computer users delegate computations to quantum servers, they wish to keep information about their calculations hidden from the servers.
The protocol of delegating a calculation while hiding information about the calculation from the server is called {\sl blind quantum computation protocol}.
Prior research on single-server's blind quantum computation protocol required users to have quantum capabilities.
Prior research on multiple-servers' blind quantum computation protocols required users to have just classical capabilities but imposed limits on the server-to-server communication.
There are no known single-server blind quantum computation protocols with a classical user and multiple-servers blind quantum computation protocols that allows servers to communicate freely with each other.
We show that the existence of these protocols is equivalence.
\end{abstract}

\section{Introduction}
Quantum computers are expected to become the next-generation computers because they can perform calculations that are considered impossible with classical computers.
For example, Shor's algorithm\cite{Shor} solves prime factorization problems in polynomial time using the quantum Fourier transform, and Grover's algorithm\cite{Grover} is recognized as the quickest unordered database search.
However, due to the sensitivity of quantum states to external noise, the physical implementation of quantum computers is hard and requires expensive technology.
As a result, quantum computers will most likely be employed as servers in cloud services rather than being owned by individual customers.
An essential concern with such cloud services is that a service provider may gain information on a calculation delegated by a user unlawfully.
As a result, a form of security is required, namely blind quantum computation protocols, which would allow users to perform calculations without revealing their contents
\cite{Previous research1,Childs,Previous research2,BFK,FK,MF protocol,Morimae hayashi,Reichardt,McKague,Sano1,Sano2}.
The inputs, outputs, and processes of these blind quantum computation protocols are encrypted.
The classical analog is known as the classical delegated computation protocols\cite{BCC1,BCC2}.
However, those are only computationally-secure.
The blind quantum computation protocol has information-theoretic security.
Thus, blind quantum computation protocols are thought to be more secure than classical delegated computation protocols.
Childs showed that a user with quantum memory and the ability to manipulate qubits, i.e., a user with a mini quantum computer, may execute blind quantum computation via quantum communication with a server equipped with a universal quantum computer\cite{Childs}.
Broadbent, Fitzsimons, and Kashefi proposed a protocol that users, who do not have quantum memory, create a specific quantum state, send it to a server, and do classical communication with the server\cite{BFK}.
Several more blind quantum computation protocols are also carried out by a user doing quantum communication with a single server\cite{FK,MF protocol,Morimae hayashi,Sano1}.
Protocols utilizing many servers have been proposed to ease the limitations on the user's abilities\cite{Reichardt,McKague}.
In these protocols, a user requires a classical computer and classical communication with multiple servers that share entangled qubits.
These protocols are useful because the user does not need to have any quantum equipment. 
However, it is vital to note that classical and quantum communication is not allowed among multiple servers.

As mentioned above, thus far, several different blind quantum computation protocols have been proposed.
However, it is uncertain if there is a single server protocol with users who only have classical capabilities and a multiple servers protocol that allows servers to freely communicate. 
The standard users are considered to have classical capabilities.
In general, servers are considered to communicate freely with each other.
Therefore, if they exist, these protocols would be the most user-friendly blind quantum computation protocols.
Our goal is to investigate the link between these protocols.

In this study, we show that if there exists a single-server blind quantum computation protocol with users who have only classical capabilities, there is a multi-server blind quantum computation protocol that enables servers to communicate freely with each other, and vice versa.
We show specifically that if the single-server protocol exists, it can be emulated with multiple servers, and that if the multiple-servers protocol exists, it can be simulated with a single server.
We further show that these simulation approaches are not affected by the particular blind quantum computation protocol configuration. 
As a result of our findings, even investigating multi-server protocols can lead to the search for blind quantum computation protocols that employ a single server with users who only have classical computational capabilities.

\section{Preliminaries}
In this section, we describe a blind quantum computation protocol.
Then, we define single server protocol with users who only have classical capabilities and a multiple-servers protocol that allows servers to freely communicate.
Note that in the following, $n$ represents the number of input bits.

\subsection{Blind quantum computation protocols}
In this subsection, we first describe blind quantum computation protocols.
A blind quantum computation protocol, first proposed by Childs\cite{Childs}, is a security feature that hides not only the input and output but also the computation algorithms from the server.
This means that when using a blind quantum computation protocol, the server does not even know what calculation the user has performed.
Naturally, quantum computation trivially includes classical computation; thus, the classical computation can also use blind quantum computation protocols as part of quantum computation.
Broadbent, Fitzsimons, and Kashefi gave the following definition for a blind quantum computation protocol\cite{BFK}.

\begin{defn}[Blindness{\cite[Definition 2]{BFK}}]
\label{defn:Blindness}
Let P be a quantum delegated computation on input X and let L(X) be any function
of the input. We say that a quantum delegated computation protocol is blind while leaking at most
L(X) if, on user's input X, for any fixed Y = L(X), the following two hold when given Y :
\begin{itemize}
 \item[1.] The distribution of the classical information obtained by server in P is independent of X.
 \item[2.]Given the distribution of classical information described in 1, the state of the quantum system
obtained by server in P is fixed and independent of X.
\end{itemize}
\end{defn}

In this paper, let the condition of definition \ref{defn:Blindness} be called {\sl blindness}, and let the protocol that satisfies the blindness be called a blind quantum computation protocol.
This definition refers to the fact that the server only gets information obtained through calculations, such as the size of the circuit, and not information that is dependent on the calculation.

\subsection{Single-server Protocol and Multi-server Protocol}
In this subsection, we define a single-server blind quantum computation protocol in which users can only use classical capabilities, and a multi-server blind quantum computation protocol in which servers can freely communicate with each other.
We will assume in the following section that honest servers have quantum computing power and malicious servers have unbounded computing power.

We first define a single-server blind quantum computation protocol in which the user has only classical capabilities.
\begin{defn}[Single-server blind quantum computation protocol with a classical user]
A user has classical computing and classical communication capabilities.
If the following user-server interaction's delegating computation protocol satisfies blindness, we define it as a single-server blind quantum computation protocol with a classical user.
The number of protocol steps $p(n)$ is the polynomial size of $n$.
\begin{description}
        \item[\textbf{Step 1.}] \textbf{Send the first message to the server}\\
        The user sends a classical polynomial-sized message $m_1$ to the server.
        \item[\textbf{Step 2.}] \textbf{Return a first message to the user}\\
        The server receives the user's message $m_1$ and performs quantum computation based on the message.
        The server transmits to the user a classical polynomial-sized message $s_1$, the content of which is determined by the server's calculation.
        \item[\textbf{Step 3.}] \textbf{Send the second message to the server}\\
        The user gets the message $s_1$ and performs classical computation based on the message.
        The user sends a classical polynomial-sized message $m_2$ to the server, the size which relies on the content of the user's calculation.\\
        $\vdots$
        \item[\textbf{Step $2i$.}] \textbf{Return an $i$-th message to the user}\\
        The server receives the user's message $m_i$ and performs quantum computation based on the message.
        The server sends a classical polynomial-sized message $s_i$, which depends on the content of the server's calculation, to the user.
        \item[\textbf{Step $2i+1$.}] \textbf{Send a $i+1$-th message to the server}\\
        The user receives the message $s_i$ and performs classical computation based on the message.
        The user sends a classical polynomial-sized message $m_{i+1}$, which depends on the content of the user's calculation, to the server.\\
        $\vdots$
        \item[\textbf{Step p(n).}] \textbf{Calculation is complete}\\
        The user receives the last message $s_l$ and obtains the result of the delegated calculation by executing a classical calculation.
        \\
       \end{description}
\end{defn}

By this definition, an honest server has quantum computing power, so obviously, a user can delegate quantum computation to it.

Next, we define a multi-server blind quantum computation protocol that allows servers to communicate with each other during computation.
We define separately when servers share entanglement with each other and when they do not.

\begin{defn}[Multiple-servers without entanglement blind quantum computation protocol that allows servers to communicate freely with each other]
A user is capable of both classical computing and classical communication.
The number of servers is polynomial-size $q(n)$.
Servers do not share quantum entanglements, and only classical communication is allowed between servers.
If the following user-server interaction's delegating computation protocol satisfies blindness, we define it as a multiple-servers without entanglement blind quantum computation protocol that allows servers to communicate freely with each other.
The number of protocol steps $p(n)$ is the polynomial size of $n$.
\begin{description}
        \item[\textbf{Step 1.}] \textbf{Send first messages to the servers}\\
        The user sends classical polynomial-sized messages to all servers.
        Let $m_{1,j}$ be the message that the user sends to the $j$-th server.
        \item[\textbf{Step 2.}] \textbf{Return first messages to the user}\\
        The $j$-th server receives the user's message $m_{1,j}$ and performs quantum computation and classical communication with other servers based on the message.
        The $j$-th server sends a classical polynomial-sized message $s_{1,j}$, which depends on the content of the server's calculation, to the user.
        \item[\textbf{Step 3.}] \textbf{Send second messages to the server}\\
        The user gets the messages $\{s_{1,j}\}_j$ and performs classical computation based on the message.
        The user sends classical polynomial-sized messages to all servers, the size of which depends on the content of the user's calculation.
        Let $m_{2,j}$ be the message that the user sends to the $j$-th server.\\
        $\vdots$
        \item[\textbf{Step $2i$.}] \textbf{Return $i$-th messages to the user}\\
        The $j$-th server gets the user's message $m_{i,j}$ and performs quantum computation classical communication with other servers based on the message.
        The $j$-th server sends a classical polynomial-sized message $s_{i,j}$, which depends on the content of the server's calculation, to the user.
        \item[\textbf{Step $2i+1$.}] \textbf{Send $i+1$-th messages to the server}\\
        The user receives the messages $\{s_{i,j}\}_j$ and performs classical computation based on the message.
        The user sends classical polynomial-sized messages to all servers, the size of which depends on the content of the user's calculation.
        Let $m_{i+1,j}$ be the message that the user sends to the $j$-th server.\\
        $\vdots$
        \item[\textbf{Step p(n).}] \textbf{Calculation is complete}\\
        The user receives the last messages $\{s_{l,j}\}_j$ from the servers and obtains a result about the delegated calculation by performing a classical calculation.
        \\
       \end{description}
\end{defn}

\begin{defn}[Multiple-servers without entanglement blind quantum computation protocol that allows servers to communicate freely with each other]
A user has classical computing and classical communication capabilities.
The number of servers is polynomial-size $q(n)$.
Servers share quantum entanglements, and servers can communicate in both classical and quantum.
If the following user-server interaction's delegating computation protocol satisfies blindness, we define it as a multiple-servers without entanglement blind quantum computation protocol that allows servers to communicate freely with each other.
The number of protocol steps $p(n)$ is the polynomial size of $n$.
\begin{description}
        \item[\textbf{Step 1.}] \textbf{Send first messages to the servers}\\
        The user sends classical polynomial-sized messages to all servers.
        Let $m_{1,j}$ be the message that the user sends to the $j$-th server.
        \item[\textbf{Step 2.}] \textbf{Return first messages to the user}\\
        The $j$-th server receives the user's message $m_{1,j}$ and performs quantum computation and classical/quantum communication with other servers based on the message.
        The $j$-th server sends a classical polynomial-sized message $s_{1,j}$, which depends on the content of the server's calculation, to the user.
        \item[\textbf{Step 3.}] \textbf{Send second messages to the server}\\
        The user gets the messages $\{s_{1,j}\}_j$ and performs classical computation based on the message.
        The user sends classical polynomial-sized messages to all servers, the size of which depends on the content of the user's calculation.
        Let $m_{2,j}$ be the message that the user sends to the $j$-th server.\\
        $\vdots$
        \item[\textbf{Step $2i$.}] \textbf{Return $i$-th messages to the user}\\
        The $j$-th server gets the user's message $m_{i,j}$ and performs quantum computation and classical/quantum communication with other servers based on the message.
        The $j$-th server sends a classical polynomial-sized message $s_{i,j}$, which depends on the content of the server's calculation, to the user.
        \item[\textbf{Step $2i+1$.}] \textbf{Send $i+1$-th messages to the server}\\
        The user receives the messages $\{s_{i,j}\}_j$ and performs classical computation based on the message.
        The user sends classical polynomial-sized messages to all servers, the size of which depends on the content of the user's calculation.
        Let $m_{i+1,j}$ be the message that the user sends to the $j$-th server.\\
        $\vdots$
        \item[\textbf{Step p(n).}] \textbf{Calculation is complete}\\
        The user receives the last messages $\{s_{l,j}\}_j$ from the servers and obtains a result about the delegated calculation by performing a classical calculation.
        \\
       \end{description}
\end{defn}

These these definitions just state that the blind protocols performed by the aforementioned processes, if they exist, will be referred to by the names provided in each definition and they do not prove the existence of these protocols.

Whether the server shares entanglement or not, the user can delegate quantum computation to the server if the server is honest because the server has quantum computation capability.
There is no requirement for actual quantum communication in the protocol with shared entanglement because quantum teleportation is conceivable by utilizing classical communication plus entanglement.

\section{Equivalence of single server and multiple server blind quantum computation protocols}
In this section, we show that if the single-server blind quantum computation protocol defined in the previous section exists, then there is a multi-server blind quantum computation protocol that allows servers to communicate with each other, and vice versa, if the multi-server blind quantum computation protocol exists, then there is the single-server blind quantum computation protocol.

\begin{thm}
\label{thm:no-entangle}
If a single-server blind quantum computation protocol with a classical user exists, then a multiple-servers without entanglement blind quantum computation protocol that allows servers to communicate freely with each other also exists.
Furthermore, if a multiple-servers without entanglement blind quantum computation protocol that allows servers to communicate freely with each other exists, then so does  a single-server blind quantum computation protocol with a classical user.
\end{thm}
\begin{proof}
We first show that if there exists a single-server blind quantum computation protocol with a classical user, then there exists a multiple-servers without entanglement blind quantum computation protocol that allows servers to communicate freely with each other.
Assume there is a single-server blind quantum computation protocol with a classical user.
The number of servers is polynomial-size $q(n)$.
The user chooses one of those servers.
This chosen server can be the first server without loss of generality.
With the following protocol, we explore the scenario when a user delegates computation to multiple servers.
It is important to note that the terms $m_i$ and $s_i$ relate to messages in the single-server blind quantum computation protocol with a classical user.
\begin{description}
        \item[\textbf{Step 1.}] \textbf{Send first messages to the servers}\\
        The user sends classical polynomial-sized messages to all servers.
        Let $m_{1,j}$ be the message that the user sends to the $j$-th server, and $m_{1,1}=m_1$ and $j\neq 1$ message $m_{1,j}$ is a meaningless string.
        \item[\textbf{Step 2.}] \textbf{Return first messages to the user}\\
        The $j$-th server receives the user's message $m_{1,j}$ and performs quantum computation and classical communication with other servers based on the message.
        The $j$-th server sends a classical polynomial-sized message $s_{1,j}$, which depends on the content of the server's calculation, to the user.
        \item[\textbf{Step 3.}] \textbf{Send second messages to the server}\\
        The user gets the message $s_{1,1}$, discards the messages from other servers, and performs classical computation based on the message.
        The user sends classical polynomial-sized messages to all servers, the size of  which depends on the content of the user's calculation.
        Let $m_{2,j}$ be the message that the user sends to the $j$-th server, and $m_{2,1}=m_2$ and $j\neq 1$ message $m_{2,j}$ is a meaningless string.\\
        $\vdots$
        \item[\textbf{Step $2i$.}] \textbf{Return $i$-th messages to the user}\\
        The $j$-th server receives the user's message $m_{i,j}$ and performs quantum computation classical communication with other servers based on the message.
        The $j$-th server sends a classical polynomial-sized message $s_{i,j}$, which depends on the content of the server's calculation, to the user.
        \item[\textbf{Step $2i+1$.}] \textbf{Send $i+1$-th messages to the server}\\
        The user receives the message $s_{i,1}$ and discards other server's messages, and performs classical computation based on the message.
        The user sends classical polynomial-sized messages to all servers, the size of which depends on the content of the user's calculation.
        Let $m_{i+1,j}$ be the message that the user sends to the $j$-th server, and $m_{i+1,1}=m_{i+1}$ and $j\neq 1$ message $m_{i+1,j}$ is a meaningless string.\\
        $\vdots$
        \item[\textbf{Step p(n).}] \textbf{Calculation is complete}\\
        The user receives the last message $s_{l,1}$ from the first server and gets a result about the delegated calculation by performing a classical calculation.
        \\
\end{description}
This protocol delegates the computation to only one server out of multiple servers.
The information gained by multiple servers during this protocol is the same as that obtained by a single server during the single-server blind quantum computation protocol with a classical user. 
If malicious servers can obtain information about the computation from this protocol, then the malicious server can also obtain information from the single-server protocol.
This contradicts the assumption.
Therefore, if there is a single-server blind quantum computation protocol with a classical user, there is a multiple-servers without entanglement blind quantum computation protocol that allows servers to communicate freely with each other.

We then show that if there exists a multiple-servers without entanglement blind quantum computation protocol that allows servers to communicate freely with each other, then there exists a single-server blind quantum computation protocol with a classical user.
Assume there is a multiple-servers without entanglement blind quantum computation protocol that allows servers to communicate freely with each other.
We consider the scene where a user delegates computation to a single server using the protocol described below.
Note that $m_{i,j}$ and $s_{i,j}$ refer to messages in the multiple-servers without entanglement blind quantum computation protocol that allows servers to communicate freely with each other.
\begin{description}
        \item[\textbf{Step 1.}] \textbf{Send the first message to the server}\\
        The user sends a classical polynomial-sized message $m_1=\{m_{1,1},\cdots,m_{1,q(n)}\}$ to the server.
        \item[\textbf{Step 2.}] \textbf{Return the first message to the user}\\
        The server receives the user's message $m_1$ and performs quantum computation based on the message.
        The server sends a classical polynomial-sized message $s_1=\{s_{1,1},\cdots,s_{1,q(n)}\}$, which depends on the content of the server's calculation, to the user.
        \item[\textbf{Step 3.}] \textbf{Send a second message to the server}\\
        The user gets the message $s_1$ and performs classical computation based on the message.
        The user sends a classical polynomial-sized message $m_2=\{m_{2,1},\cdots,m_{2,q(n)}\}$, which depends on the content of the user's calculation, to the server.\\
        $\vdots$
        \item[\textbf{Step $2i$.}] \textbf{Return a $i$-th message to the user}\\
        The server receives the user's message $m_i$ and performs quantum computation based on the message.
        The server sends a classical polynomial-sized message $s_i=\{s_{i,1},\cdots,s_{i,q(n)}\}$, which depends on the content of the server's calculation, to the user.
        \item[\textbf{Step $2i+1$.}] \textbf{Send a $i+1$-th message to the server}\\
        The user receives the message $s_i$ and performs classical computation based on the message.
        The user sends a classical polynomial-sized message $m_{i+1}=\{m_{i+1,1},\cdots,m_{i+1,q(n)}\}$, which depends on the content of the user's calculation, to the server.\\
        $\vdots$
        \item[\textbf{Step p(n).}] \textbf{Calculation is complete}\\
        The user receives the last message $s_l$ from the server and gets a result about the delegated calculation by performing a classical calculation.
        \\
\end{description}
This protocol may be thought of as a single server simulation of the multiple-servers without entanglement blind quantum computation protocol that allows servers to communicate freely with each other.
Malicious servers can do classical communication during computation in the multiple-servers without entanglement blind quantum computation protocol that allows servers to communicate freely with each other.
In other words, malicious servers might transmit all user messages to a single server and calculate them alone on that server.
Since the malicious server has unbounded computing power, there is no difference in computing power whether all calculations are alone on one server or multiple servers.
The multiple-server blind quantum computation protocol satisfies blindness to such attacks by malicious servers by assumption.
If the malicious single server can get calculation information from the aforementioned single-server protocol, then malicious servers can also get calculation information from the multiple-server protocol.
This contradicts the assumption.
Therefore, if a multiple-servers without entanglement blind quantum computation protocol that allows servers to communicate freely with each other exists, so does a single-server blind quantum computation protocol with a classical user.
\end{proof}

\begin{thm}
If a single-server blind quantum computation protocol with a classical user exists, then a multiple-servers without entanglement blind quantum computation protocol that allows servers to communicate freely with each other also exists.
Furthermore, if a multiple-servers without entanglement blind quantum computation protocol that allows servers to communicate freely with each other exists, then so does a single-server blind quantum computation protocol with a classical user.
\end{thm}
\begin{proof}
The proof is the same as in Theorem \ref{thm:no-entangle}: If a single-server blind quantum computation protocol with a classical user exists, then a multiple-server blind quantum computation protocol with entanglement also exists.

We show that if there exists a multiple-servers without entanglement blind quantum computation protocol that allows servers to communicate freely with each other, then there exists a single-server blind quantum computation protocol with a classical user.
Assume a multiple-servers without entanglement blind quantum computation protocol that allows servers to communicate freely with each other exists.
We consider the case where a user delegates computation to a single server using the protocol described below.
The number of servers is polynomial-size $q(n)$.
Note that $m_{i,j}$ and $s_{i,j}$ refer to messages in the multiple-servers without entanglement blind quantum computation protocol that allows servers to communicate freely with each other.
\begin{description}
        \item[\textbf{Step 1.}] \textbf{Send the first message to the server}\\
        The user sends a classical polynomial-sized message $m_1=\{m_{1,1},\cdots,m_{1,q(n)}\}$ to the server.
        \item[\textbf{Step 2.}] \textbf{Return the first message to the user}\\
        The server receives the user's message $m_1$ and performs quantum computation based on the message.
        The server sends a classical polynomial-sized message $s_1=\{s_{1,1},\cdots,s_{1,q(n)}\}$, which depends on the content of the server's calculation, to the user.
        \item[\textbf{Step 3.}] \textbf{Send a second message to the server}\\
        The user gets the message $s_1$ and performs classical computation based on the message.
        The user sends a classical polynomial-sized message $m_2=\{m_{2,1},\cdots,m_{2,q(n)}\}$, which depends on the content of the user's calculation, to the server.\\
        $\vdots$
        \item[\textbf{Step $2i$.}] \textbf{Return a $i$-th message to the user}\\
        The server receives the user's message $m_i$ and performs quantum computation based on the message.
        The server sends a classical polynomial-sized message $s_i=\{s_{i,1},\cdots,s_{i,q(n)}\}$, which depends on the content of the server's calculation, to the user.
        \item[\textbf{Step $2i+1$.}] \textbf{Send a $i+1$-th message to the server}\\
        The user receives the message $s_i$ and performs classical computation based on the message.
        The user sends a classical polynomial-sized message $m_{i+1}=\{m_{i+1,1},\cdots,m_{i+1,q(n)}\}$, which depends on the content of the user's calculation, to the server.\\
        $\vdots$
        \item[\textbf{Step p(n).}] \textbf{Calculation is complete}\\
        The user receives the last message $s_l$ from the server and gets a result about the delegated calculation by performing a classical calculation.
        \\
\end{description}
A single server can also easily prepare entanglement, making such protocols feasible.
This protocol can be interpreted as a simulation by a single server of the multiple-servers without entanglement blind quantum computation protocol that allows servers to communicate freely with each other.
In the multiple-servers without entanglement blind quantum computation protocol that allows servers to communicate freely with each other, malicious servers can do classical/quantum communication during computation.
However, we are not required to consider the quantum communication that the malicious servers do, because the quantum states that what each malicious server can prepare can also be prepared by other malicious servers.
In other words, malicious servers might transmit all user messages to a single server and calculate them alone on the server.
Since the malicious server has unbounded computing power, it makes no difference in computing power whether all computations are performed on a single server or multiple servers.
The multiple-server blind quantum computation protocol satisfies blindness to such attacks by malicious servers by assumption.
If the malicious single server can get calculation information from the aforementioned single-server protocol, then malicious servers can also get calculation information from the multiple-server protocol.
This contradicts the assumption.
Therefore, if a multiple-servers with entanglement blind quantum computation protocol that allows servers to communicate freely with each other exists, so does a single-server blind quantum computation protocol with a classical user.
\end{proof}

\section{Discussion}
In this research, we have defined a single-server blind quantum computation protocol with a classical user, a multiple-servers without entanglement blind quantum computation protocol that allows servers to communicate freely with each other, and a multiple-servers without entanglement blind quantum computation protocol that allows servers to communicate freely with each other, and have proved the equivalence of the existence of these protocols.
It is not known if a single-server blind quantum computation protocol with a classical user exists\cite{Fitzsimons review,morimae hitei,aronson hitei}.
As a result, it is a significant open problem.
Multi-server blind protocols are helpful but have received little attention.
Our results imply that investigating multi-server blind protocols can reveal the existence of a single-server blind quantum computation protocol with a classical user.

\section*{Acknowledgment}
We would like to thank Takayuki Miyadera for the many helpful comments, and we are grateful to Kazuki Yamaga for his important advice.
This work was supported by JST SPRING, Grant Number JPMJSP2110.

\end{document}